\documentclass{iopconfser}
\usepackage[compress]{cite}
\usepackage{amsmath, amsthm, amssymb}
\usepackage{amsmath, amsthm, amssymb}
\usepackage{graphicx}
\usepackage{enumitem}
\usepackage{multirow}

\begin{document}

\title{Can Relativistic Effects explain Galactic Dynamics without Dark Matter?}

\author{L. Filipe O. Costa$^{1}$, José Natário$^{2}$}

\affil{$^1$Centro de Matemática (CMAT), University of Minho, Portugal}
\affil{$^2$CAMGSD, Instituto Superior Técnico, Portugal}

\email{filipecosta@cmat.uminho.pt}

\begin{abstract}
%Assuming only stationarity of the field, 
We show that, contrary to some recent claims, relativistic effects cannot mimic dark matter in the galactic rotation curves and gravitational lensing.
\end{abstract}

\section{Missing mass problem in the framework of weak field approximations}
%According to well established approximations to General Relativity (GR), namely the post-Newtonian (PN) and post-Minskowskian expansions, the galactic rotation curves and gravitational lensing cannot be explained based only on the visible baryonic matter. It is also clear in these frameworks that relativistic effects cannot resolve it: consider geometrized units where $c=G=1$.
According to the conventional understanding of gravity laws, namely General Relativity (GR), 
%General Relativity (GR) 
the galactic rotation curves and gravitational lensing cannot be explained based only on the visible baryonic matter \cite{Ciotti:2022inn,GlampedakisJones_Pitfalls,LasenbyHobsonBarker}. In the framework of the post-Newtonian (PN) expansion of GR, it is also clear that relativistic effects cannot be the answer: consider geometrized units where $c=G=1$. For stars in a galaxy, $v\lesssim10^{-3}$; it follows from e.g. Eq. (8.16) of \cite{WillPoissonBook}, or Eq. (7.17) of \cite{Damour:1990pi}, that % $d^{2}\vec{x}/dt^{2}=\vec{G}_{{\rm N}}+O(|\vec{G}_{{\rm N}}|10^{-6})$, $\vec{G}_{{\rm N}}\equiv$ Newtonian field 
\begin{center}
\begin{tabular}{cl}
\multirow{2}{*}{${\displaystyle \frac{d^{2}\vec{x}}{dt^{2}}=\vec{G}_{{\rm N}}+O(|\vec{G}_{{\rm N}}|10^{-6})} \, ; $} & \multirow{2}{*}{~~~~~~$\vec{G}_{{\rm N}}\equiv$ Newtonian field,}\tabularnewline
 & \tabularnewline
\end{tabular}
\end{center}
showing that relativistic corrections are \textit{one million} times smaller than needed to impact rotation curves.
It has however been argued, in recent literature (e.g. \cite{RuggieroBG,RuggieroGalatic,AstesianoRuggieroPRDII,BG,Crosta2018}), that the PN expansion might overlook %neglect from the outset% 
gravitomagnetic and/or non-linear GR effects that could mimic dark matter. In what follows we address the problem without resorting to approximations. 

\section{Exact theory \label{exacttheory}}
The line element of a stationary spacetime can be written in the form 
\begin{equation}
ds^{2}=-e^{2\Phi}(dt-\mathcal{A}_{i}dx^{i})^{2}+h_{ij}dx^{i}dx^{j}\ ,\label{eq:StatMetric}
\end{equation}
with $\Phi\equiv\Phi(x^{j})$, $\mathcal{A}_{i}\equiv\mathcal{A}_{i}(x^{j})$,
and $h_{ij}\equiv h(x^{k})_{ij}$.

The space components of the geodesic equation $d^{2}x^{\alpha}/d\lambda^{2}=-\Gamma_{\mu\nu}^{\alpha}(dx^{\mu}/d\lambda)(dx^{\nu}/d\lambda)$ read
\begin{equation}
\frac{d^{2}x^{i}}{d\lambda^{2}}+\Gamma(h)_{jk}^{i}\frac{dx^{j}}{d\lambda}\frac{dx^{k}}{d\lambda}\equiv \frac{\tilde{D}}{d\lambda} \left[\frac{d x^{i}}{d\lambda}\right]=\nu^2\left[\vec{G}+\vec{v}\times\vec{H}\right]^{i},\label{eq:QMGeoGeneral}
\end{equation}
where $\nu\equiv-u_{\alpha}dx^{\alpha}/d\lambda$, $u^{\alpha}=e^{-\Phi}\partial_{t}^{\alpha}\equiv$ 4-velocity of the static observers $O(u)$, $\vec{v}=(1/\nu)d\vec{x}/d\lambda$ is the spatial velocity of the particle relative to $O(u)$, $\Gamma(h)_{jk}^{i}\equiv h^{il}\left(h_{lj,k}+h_{lk,j}-h_{jk,l}\right)/2$ are the Christoffel symbols of the space metric $h_{ij}$, $\tilde{D}/d\tau\equiv$ covariant derivative with respect to $h_{ij}$ and
\begin{equation}
G_{i}=-\Phi_{,i}\ ;\qquad\ H^{i}=e^{\Phi}\epsilon^{ijk}\mathcal{A}_{k,j}\quad(\epsilon_{ijk}\equiv\sqrt{h}[ijk])\label{eq:GEM1forms}
\end{equation}
are the ``gravitoelectric'' and ``gravitomagnetic'' fields, playing in Eq. \eqref{eq:QMGeoGeneral} roles analogous to the electric and
magnetic fields in the Lorentz force equation $D\vec{U}/d\tau=(q\gamma/m)[\vec{E}+\vec{v}\times\vec{B}]$ \cite{CostaNatarioGalactic2023}. Eq. \eqref{eq:QMGeoGeneral} splits into 
\stepcounter{equation}

\begin{tabular}{lllllllll}
~~~~~~~~~~~~~~\raisebox{3ex}{} Time-like geodesic &  &  &  & Null geodesic &  &  &  & \tabularnewline
~~~~~~~~~~~~~~\raisebox{2ex}{${\displaystyle \frac{\tilde{D}\vec{U}}{d\tau}=\gamma^{2}\left[\vec{G}+\vec{v}\times\vec{H}\right]}$}\raisebox{5.5ex}{} & \raisebox{2ex}{{\small{}~($v<1)$}} &  &  & \raisebox{2ex}{${\displaystyle \frac{\tilde{D}\vec{k}}{d\lambda}=\nu^{2}\left[\vec{G}+\vec{v}\times\vec{H}\right]}$} & \raisebox{2ex}{{\small{}~($v=1)$}} &  &  & ~~~~~\raisebox{2ex}{(\theequation)}\tabularnewline
\end{tabular} \\
where $U^{\alpha}\equiv dx^{\alpha}/d\tau$ is the 4-velocity, $d\tau\equiv$ proper time, and $k^{\alpha}=dx^{\alpha}/d\lambda$  a null vector ($k^{\alpha} k_{\alpha}=0$).

The first of Eqs. (\theequation) tells us that, if relativistic effects were to explain the galactic rotation curves, it would have to be through $\vec{H}$, or non-linear contributions to $\vec{G}$.

\subsection{Gravitomagnetism not the culprit---gravitational lensing \label{GMnotculprit}}
\subsubsection*{$\vec{H}$ cannot originate the observed gravitational lensing}
Nearly perfect Einstein rings, generated by galactic gravitational lenses, have been observed, e.g. B1938+666 \cite{King_first_Einstein_Ring,Lagattuta_Einstein_ring,Petters_2001singularity},
J1148+1930 (``Cosmic Horseshoe'') \cite{Belokurov_Horseshoe_2007,Spiniello_horseshoe_2011,Bellagamba_horseshoe_2011,Schuldt_et_al_horseshoe_2019,Cheng_et_al_horseshoe_2019}, COSMOS-Web ring \cite{Cosmos-Web_ring}, JWST-ER1 \cite{JWST_ER1}. They are formed when the lens is nearly spherical, and the light source, lens, and observer are nearly aligned. The majority of the lens' mass is estimated to be dark matter \cite{Spiniello_horseshoe_2011,Bellagamba_horseshoe_2011,Schuldt_et_al_horseshoe_2019,Cheng_et_al_horseshoe_2019,Lagattuta_Einstein_ring,Cosmos-Web_ring}, and is consistent with dark matter halos roughly spherical \cite{Schuldt_et_al_horseshoe_2019,Durkalec}.
The gravitomagnetic field $\vec{H}$ cannot produce these effects: assuming stationarity, axisymmetry, and reflection symmetry about the equatorial plane, we have $\mathcal{A}_{i}dx^{i}=\mathcal{A}_{\phi}d\phi$ and, in the equatorial plane, $\mathcal{A}_{\phi,i}=\mathcal{A}_{\phi,r}\delta_{i}^{r}$, implying, via Eq. (3), that $\vec{H}$ is axisymmetric and orthogonal to the equatorial plane. 
%Moreover, for a finite rotating body, $\vec{H}$ is dipole-like in the far field \cite{HarrisAnalogy1991}.
Such a field is moreover (for a finite rotating body) dipole-like at large distances \cite{HarrisAnalogy1991}. 
Therefore, as displayed in Figs. 1 and 2 of \cite{CostaNatarioGalactic2023}, in the equatorial plane, the ``force'' $\vec{v}\times\vec{H}$ deflects light rays on both sides of the body in the same direction; the same occurs at the poles, in opposite direction. This does not create the symmetric convergence of rays along the optical axis, actually acting against the formation of the ring (cf. \cite{Perlick_Lensing_2004,RauchBlandford_Lensing_Kerr,Bozza_Kerr_Equat,Sereno_De_Luca_Kerr,Sereno_Weak}).

\begin{tabular}{c}
\includegraphics[width=0.7\textwidth]{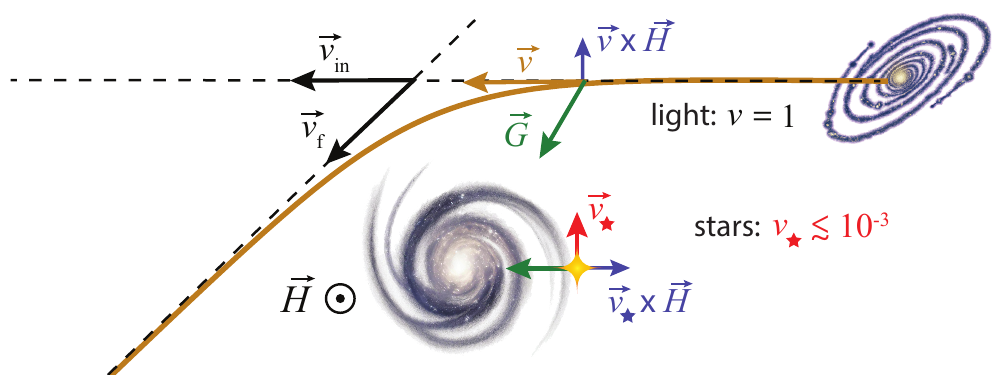}\tabularnewline
\end{tabular}
\subsubsection*{Any $\vec{H}$ large enough to impact rotation curves yields unacceptable lensing effects} 
%\subsubsection*{Any $\vec{H}$ large enough to impact rotation curves, yields unacceptable lensing effects}
%\begin{figure}[h!]
%\end{figure}
%\noindent\includegraphics[width=0.7\textwidth]{LensingConstraint}
For stars in a galaxy $v\equiv v_{\star}\lesssim 10^{-3}$ \cite{Sofue2000,Eilers2019}; thus, in order for the gravitomagnetic force $\vec{v}_{\star}\times\vec{H}$
to have an impact on rotation curves, one needs $|\vec{H}|\sim10^{3}|\vec{G}|$; by Eqs. (\theequation) this would imply, for light (since $v=1$), $|\vec{v}\times\vec{H}|\sim10^{3}|\vec{G}|$, i.e., a gravitational lens with gravitomagnetic force 3 orders of magnitude larger than the Newtonian force, and therefore (by e.g Eq. (28) of  \cite{CostaNatarioGalactic2023}) bending angles \emph{orders of magnitude larger} than observed.
% Such a large $\vec{H}$ is moreover impossible for a rotating body: $H/G\sim v_{{\rm rot}}R/r<1$;  would require singularities.

\subsection{Non-linear relativistic contributions \textit{reduce} attraction}
The observed lensing effects thus constrain the magnitude of $\vec{H}$ to be, at most, within the same order of magnitude of $\vec{G}$. This constrains the geodesic equation for stars in a galaxy to $\tilde{D}\vec{U}/d\tau=\vec{G}+O(<10^{-3} |\vec{G}|)$, and their velocities to $v_{\star}=\sqrt{rG_{r}}+O(<10^{-6})$ \cite{CostaNatarioGalactic2023}. Therefore, it remains only to clarify whether non-linear GR effects can amplify $\vec{G}$ in order to produce an attractive effect able to sustain the rotation curves without dark matter. This is however ruled out by the field equations for $\vec{G}$ and $\vec{H}$
\begin{equation}
\tilde{\nabla}\cdot\vec{G}=-4\pi(2\rho+T_{\ \alpha}^{\alpha})+{\vec{G}}^{2}+\frac{1}{2}{\vec{H}}^{2}\ ;\qquad\tilde{\nabla}\times\vec{H}=-16\pi\vec{J}+2\vec{G}\times\vec{H}\ ;\qquad\tilde{\nabla}\times\vec{G}=0\ ;\qquad\tilde{\nabla}\cdot\vec{H}=-\vec{G}\cdot\vec{H}\label{eq:FieldEqs}
\end{equation}
[here $\tilde{\nabla}$ denotes covariant differentiation with respect
to the spatial metric $h_{ij}$, and the equations for $\tilde{\nabla}\cdot\vec{G}$ and $\tilde{\nabla}\times\vec{H}$
are, respectively, the time-time and time-space projections, with respect to $u^{\alpha}=e^{-\Phi}\delta_{0}^{\alpha}$, of the Einstein field equations $R_{\alpha\beta}=8\pi(T_{\alpha\beta}^{\ }-\frac{1}{2}g_{\alpha\beta}^{\ }T_{\ \gamma}^{\gamma})$, with $\rho\equiv T^{\alpha\beta}u_{\alpha}u_{\beta}$ and $J^{\alpha}\equiv-T^{\alpha\beta}u_{\beta}$ \cite{Analogies}. The equations for $\tilde{\nabla}\cdot\vec{H}$ and $\tilde{\nabla}\times\vec{G}$ are identities that follow from \eqref{eq:GEM1forms}]:

\begin{itemize}[itemsep=0pt, parsep=4pt, topsep=0pt, partopsep=0pt]
\item the non-linear terms ${\vec{G}}^{2}$ and ${\vec{H}}^{2}/2$ act as effective \textit{negative} ``energy'' sources for $\vec{G}$, countering the attractive effect of \textrm{$2\rho+T_{\ \alpha}^{\alpha}$} $\Rightarrow$ can only \textit{aggravate} the missing mass problem.
\item This is manifest already at first post-Newtonian order (1PN), see Sec. V.A of  \cite{CostaNatarioGalactic2023} \\
(effect anyway negligible in any realistic galactic model).
\end{itemize}

\section{Conclusion}
We have demonstrated that %, contrary to some recent claims, 
relativistic effects cannot resolve the missing mass problem:
gravitational lensing rules out the gravitomagnetic field as a player, and non-linear effects can only aggravate the need for dark matter. The conclusion assumes only stationarity of the field and is otherwise general.

\section*{Acknowledgments}
LFC acknowledges the support of CMAT (Centro de Matemática da Universidade do Minho) through FCT/Portugal projects UID/00013:Centro de Matemática da Universidade do Minho (CMAT/UM) and UID/00013/2025. JN was partially supported by FCT/Portugal through CAMGSD, IST-ID (projects UIDB/04459/2020 and UIDP/04459/2020) and project 2024.04456.CERN, and also by the H2020-MSCA-2022-SE project EinsteinWaves, GA no. 101131233.

\bibliography{Ref}

\providecommand{\newblock}{}
\begin{thebibliography}{10}
\expandafter\ifx\csname url\endcsname\relax
  \def\url#1{{\tt #1}}\fi
\expandafter\ifx\csname urlprefix\endcsname\relax\def\urlprefix{URL }\fi
\providecommand{\eprint}[2][]{\url{#2}}

\bibitem{Ciotti:2022inn}
Ciotti L 2022 {\em Astrophys. J.\/} {\bf 936} 180 (\textit{Preprint}
  \eprint{2207.09736})

\bibitem{GlampedakisJones_Pitfalls}
Glampedakis K and Jones D~I 2023 {\em Class. Quant. Grav.\/} {\bf 40} 147001
  (\textit{Preprint} \eprint{2303.16679})

\bibitem{LasenbyHobsonBarker}
Lasenby A~N, Hobson M~P and Barker W~E~V 2023 {\em Class. Quant. Grav.\/} {\bf
  40} 215014 (\textit{Preprint} \eprint{2303.06115})

\bibitem{WillPoissonBook}
{Poisson} E and {Will} C~M 2014 {\em {Gravity: Newtonian, Post-Newtonian,
  Relativistic}\/} (Cambridge, UK: Cambridge University Press)

\bibitem{Damour:1990pi}
Damour T, Soffel M and Xu C 1991 {\em Phys. Rev. D\/} {\bf 43} 3273--3307

\bibitem{RuggieroBG}
Astesiano D and Ruggiero M~L 2022 {\em Phys. Rev. D\/} {\bf 106} 044061
  (\textit{Preprint} \eprint{2205.03091})

\bibitem{RuggieroGalatic}
Ruggiero M~L, Ortolan A and Speake C~C 2022 {\em Class. Quant. Grav.\/} {\bf
  39} 225015 (\textit{Preprint} \eprint{2112.08290})

\bibitem{AstesianoRuggieroPRDII}
Astesiano D and Ruggiero M~L 2022 {\em Phys. Rev. D\/} {\bf 106} L121501
  (\textit{Preprint} \eprint{2211.11815})

\bibitem{BG}
Balasin H and Grumiller D 2008 {\em Int. J. Mod. Phys. D\/} {\bf 17} 475--488
  (\textit{Preprint} \eprint{astro-ph/0602519})

\bibitem{Crosta2018}
Crosta M, Giammaria M, Lattanzi M~G and Poggio E 2020 {\em Mon. Not. Roy.
  Astron. Soc.\/} {\bf 496} 2107--2122 (\textit{Preprint} \eprint{1810.04445})

\bibitem{CostaNatarioGalactic2023}
Costa L~F~O and Nat\'ario J 2024 {\em Phys. Rev. D\/} {\bf 110} 064056
  (\textit{Preprint} \eprint{2312.12302})

\bibitem{King_first_Einstein_Ring}
King L~J {\em et~al.\/} 1998 {\em Mon. Not. Roy. Astron. Soc.\/} {\bf 295} L41
  (\textit{Preprint} \eprint{astro-ph/9710171})

\bibitem{Lagattuta_Einstein_ring}
{Lagattuta} D~J, {Vegetti} S, {Fassnacht} C~D, {Auger} M~W, {Koopmans} L~V~E
  and {McKean} J~P 2012 {\em Mon. Not. Roy. Astron. Soc.\/} {\bf 424}
  2800--2810 (\textit{Preprint} \eprint{1206.1681})

\bibitem{Petters_2001singularity}
Petters A, Levine H and Wambsganss J 2001 {\em Singularity Theory and
  Gravitational Lensing\/} Progress in Mathematical Physics (Birkh{\"a}user
  Boston)

\bibitem{Belokurov_Horseshoe_2007}
Belokurov V {\em et~al.\/} 2007 {\em Astrophys. J. Lett.\/} {\bf 671} L9
  (\textit{Preprint} \eprint{0706.2326})

\bibitem{Spiniello_horseshoe_2011}
{Spiniello} C, {Koopmans} L~V~E, {Trager} S~C, {Czoske} O and {Treu} T 2011
  {\em Monthly Notices of the Royal Astronomical Society\/} {\bf 417}
  3000--3009 (\textit{Preprint} \eprint{1103.4773})

\bibitem{Bellagamba_horseshoe_2011}
Bellagamba F, Tessore N and Metcalf R~B 2017 {\em Mon. Not. Roy. Astron.
  Soc.\/} {\bf 464} 4823--4834 (\textit{Preprint} \eprint{1610.06003})

\bibitem{Schuldt_et_al_horseshoe_2019}
{Schuldt} S, {Chiriv{\`\i}} G, {Suyu} S~H, {Y{\i}ld{\i}r{\i}m} A, {Sonnenfeld}
  A, {Halkola} A and {Lewis} G~F 2019 {\em Astronomy \& Astrophysics\/} {\bf
  631} A40 (\textit{Preprint} \eprint{1901.02896})

\bibitem{Cheng_et_al_horseshoe_2019}
{Cheng} J, {Wiesner} M~P, {Peng} E~H, {Cui} W, {Peterson} J~R and {Li} G 2019
  {\em Astrophys. J.\/} {\bf 872} 185

\bibitem{Cosmos-Web_ring}
Mercier W {\em et~al.\/} 2024 {\em {Astronomy and Astrophysics}\/} {\bf 687}
  A61 (\textit{Preprint} \eprint{2309.15986})

\bibitem{JWST_ER1}
{van Dokkum} P, {Brammer} G, {Wang} B, {Leja} J and {Conroy} C 2024 {\em Nature
  Astronomy\/} {\bf 8} 119--125 (\textit{Preprint} \eprint{2309.07969})

\bibitem{Durkalec}
Durkalec A, Pollo A and Abbas U 2024 {\em Astrophys. J.\/} {\bf 966} 73
  (\textit{Preprint} \eprint{2404.05030})

\bibitem{HarrisAnalogy1991}
Harris E~G 1991 {\em American Journal of Physics\/} {\bf 59} 421--425

\bibitem{Perlick_Lensing_2004}
Perlick V 2004 {\em Living Reviews in Relativity\/} {\bf 7} 9

\bibitem{RauchBlandford_Lensing_Kerr}
{Rauch} K~P and {Blandford} R~D 1994 {\em The Astrophysical Journal\/} {\bf
  421} 46

\bibitem{Bozza_Kerr_Equat}
Bozza V 2003 {\em Phys. Rev. D\/} {\bf 67} 103006 (\textit{Preprint}
  \eprint{gr-qc/0210109})

\bibitem{Sereno_De_Luca_Kerr}
Sereno M and De~Luca F 2008 {\em Phys. Rev. D\/} {\bf 78} 023008
  (\textit{Preprint} \eprint{0710.5923})

\bibitem{Sereno_Weak}
Sereno M 2003 {\em Mon. Not. Roy. Astron. Soc.\/} {\bf 344} 942
  (\textit{Preprint} \eprint{astro-ph/0307243})

\bibitem{Sofue2000}
Sofue Y and Rubin V 2001 {\em Ann. Rev. Astron. Astrophys.\/} {\bf 39} 137--174
  (\textit{Preprint} \eprint{astro-ph/0010594})

\bibitem{Eilers2019}
{Eilers} A~C, {Hogg} D~W, {Rix} H~W and {Ness} M~K 2019 {\em The Astrophysical
  Journal\/} {\bf 871} 120 (\textit{Preprint} \eprint{1810.09466})

\bibitem{Analogies}
Costa L~F~O and Nat{\'a}rio J 2014 {\em General Relativity and Gravitation\/}
  {\bf 46} 1792 (\textit{Preprint} \eprint{1207.0465})

\end{thebibliography}

\end{document}